\documentclass[11pt,a4wide]{article}
\usepackage{newlfont}
\usepackage{amsmath}
\usepackage{graphics}
\usepackage{float}
\usepackage{graphicx}
\usepackage{epsfig}
\usepackage{multirow}
\usepackage{multicol}
\usepackage{verbatim}
\usepackage{pdflscape}
\usepackage{cite}
\usepackage{array}
\begin{document}
	
	\title{Mass Spectrum, Radii, and Radiative Decay Widths of Toponium}
	\author{Nosheen Akbar\thanks{e mail: nosheenakbar@cuilahore.edu.pk}, ~Ishrat Asghar$^\dag$, Zaki Ahmad$^{\ddag \S}$ \\
		\textit{$\ast$Department of Physics, COMSATS University Islamabad, Lahore Campus, Pakistan}\\
		\textit{$\dag$Department of Physics, University of Education, Faisalabad Campus, Pakistan}\\
		\textit{$\ddag$Department of Physics, Govt. College Women University, Sialkot, Pakistan}\\
		\textit{$\S$Centre for High Energy Physics, Univeristy of the Punjab, Lahore, Pakistan}}
	\date{}
	\maketitle
	
	\section*{Abstract}
	In this work, radial Schrodinger equation with a non-relativistic quark potential model (NRQPM) is solved numerically by employing the shooting method. Calculated numerical wave functions (or solutions) are used to compute the masses, root mean square (RMS) radii, $E1$ and $M1$ radiative transitions, and branching ratios of $S, P, D$ and $F$ states of toponium mesons ($t\overline{t}$). Calculated results are compared with recently available theoretical data. This work will be helpful for experimentalists in gaining a deeper understanding of toponium states.
	
	\section{Introduction}
A pseudoscalar state($\eta_t$) of toponium ($t\bar{t}$) has been observed recently by CMS collaboration at CERN \cite{CMS-2025} in pp-collision with $m_{t\bar{t}} < 360 GeV$. Earlier, quantum entanglement in toponium with $340 < m_{t\bar{t}} <380$ GeV is observed by ATLAS collaboration \cite{ATLAS24}. Quantum entanglement in pair of top quarks is also confirmed by CMS collaboration \cite{CMS24} in pp-collision at $\sqrt{s}= 13$ TeV. These experimental observations attract the theorists to investigate the toponium mesons. Recently, in Ref.\cite{2025}, Bethe-Salpeter equation (BSE) is used to compute the mass spectrum, two photon, two gluon, and di-leptonic decays of the $\eta_t$ state by taking the mass of top quark equal to 172.7 GeV. 
	
Investigation of the toponium mesons has remained a topic of continuous interest since decades. Physicists did theoretical work on toponium mesons to predict their properties. But, they considered the top quark much lighter than the experimental findings as reported in Refs. \cite{CDF,ATLAS,CMS,Tevatron_and_LHC,2024,2023}. In Ref.\cite{1} toponium spectroscopy and leptonic widths are calculated through the coulomb-linear potential model for mass of top quark ($m_t=21\textrm{GeV}$ and $24 \textrm{GeV}$). An interquark potential with relativistic corrections is used in Ref.\cite{2} to predict the masses of 1S-4S and 1P-2P states of toponium mesons for top quark mass ($m_t=40\textrm{GeV}$ and $45 \textrm{GeV}$). In Ref.\cite{2}, E1 radiative transitions of 2S-4S states are also predicted at $m_t=40 \textrm{GeV}$. A non-relativistic potential with QCD vacuum polarization correction is used in Ref.\cite{3} to calculate the masses of 1S, 2S, 1P and 2P states of toponium mesons for a range of top quark mass ($30-50\textrm{GeV}$). The masses of nS-states(n=1,2,3,4,5,6) of toponium mesons are calculated in Ref.\cite{4} by solving the Schrodinger wave equation for energy-dependent potential with $20\textrm{GeV}<m_t<60\textrm{GeV}$. Decays of toponium are studied in Refs. \cite{KOPP,Ya} by assuming a very small value for mass in comparison to the recent experimental measurements. The CDF II experiment at Tevatron measure the mass of top quark ($m_t=172.85\pm0.71(stat.)\pm0.85(syst.)\textrm{GeV}/c^2$) in $p\bar{p}$ collision at $\sqrt{s}=1.96\textrm{TeV}$ \cite{CDF}. The top quark mass measured by ATLAS experiment at LHC is $m_t=172.99\pm0.41(stat.)\pm0.74(syst.)\textrm{GeV}/c^2$ at $\sqrt{s}=8\textrm{TeV}$\cite{ATLAS}. The CMS experiment present $m_t=172.44\pm0.13(stat.)\pm0.47(syst.)\textrm{GeV}/c^2$ at $\sqrt{s}=7$ and $8\textrm{TeV}$\cite{CMS}. In Ref.\cite{Tevatron_and_LHC}, a precise value of top quark mass is derived from the combination of CDF, DO, ATLAS and CMS is $m_t=173.34\pm0.76\textrm{GeV}$. In Ref.\cite{2024}, mass of top quark is suggested equal to $172.52 \pm 0.33 \textrm{GeV}$ by considering the top quark mass measurements by the ATLAS and CMS experiments in proton-proton collisions at $\sqrt{s}=7$ and 8 TeV. The mass of top quark $m_t = 174.41 \pm 0.39 (stat.) \pm 0.66 (syst.) \pm 0.25 \textrm{GeV}$ is measured by ATLAS collaboration at $\sqrt{s}=13 \textrm{GeV}$ \cite{2023}. The CMS experiment present $m_t=172.44\pm0.13(stat.)\pm0.47(syst.)\textrm{GeV}/c^2$ at $\sqrt{s}=7$ and $8\textrm{TeV}$\cite{CMS}. In the present work, mass of top quark is considered equal to $172.42\textrm{GeV}$ to calculate the properties of toponium mesons.
	
Different potential models have been developed successfully for the finding of spectrum of quarkonia: charmonium and bottomonium. In this work, a non-relativistic quark potential model is used to find the masses and wave functions of $t\bar{t}$ mesons. The spin-spin and spin-angular momentum interactions are incorporated in the columbic plus linear potential. This NRQPM has been used successfully to study the different properties of $b\bar{b}$ mesons\cite{6}. In present work, a comprehensive study of spectrum of $t\bar{t}$ is presented. Mass of $1S$ state of toponium meson ($\eta_t(1^1S_0)$) is found as $342.867$GeV which agrees with the mass range observed in recent CMS observation\cite{CMS-2025}. The wave functions calculated through NRQPM are used to find the 1)-RMS radii of toponium mesons and 2)- the radiative transitions of higher states of $t\overline{t}$ mesons upto $n L= 6 S, 3 P, 3 D, 1 F$. \\
	
The paper is organized as follows. In section 2, the potential model is described that is used to calculate the masses and wave functions of $t\overline{t}$ mesons. E1 and M1 radiative transitions of toponium mesons are calculated in Sec. 3. Results of our computations are discussed in Sec. 4; while the concluding remarks are given in Section 5.
	
\section{Potential model for toponium mesons}
	
To find the mass spectrum and wave functions of $t\overline{t}$, following non-relativistic potential model \cite{Godfrey05} is used as:
\begin{eqnarray}\label{eq:potenitalmodel}
	V_{t\bar{t}}(r) &=&-\frac{4\alpha _{s}}{3r}+br+\frac{32\pi \alpha _{s}}{
			9 m_{t} m_{\bar{t}}}(\frac{\sigma }{\sqrt{\pi }})^{3}e^{-\sigma ^{2}r^{2}}%
		\mathbf{S}_{t}.\mathbf{S}_{\bar{t}}  \notag \\
		&&+\frac{1}{m_{t} m_{\bar{t}}}[(\frac{2\alpha _{s}}{r^{3}}-\frac{b}{2r})%
		\mathbf{L}.\mathbf{S}+\frac{4\alpha _{s}}{r^{3}}T],
	\end{eqnarray}
	where $m_{t}$ is the mass of top quark, $m_{\overline{t}}$ is the mass of anti top quark, $\alpha _{s}$ is the quantum chromodynamics (QCD) running coupling constant and $b$ is the string tension. In equation (\ref{eq:potenitalmodel}) the columbic term, spin orbit term (due to the small distance between $t$ and $\overline{t}$), and tensor term are the result of one gluon exchange process. The spin orbit term (due to large distance between $t$ and $\overline{t}$) is a effect of Lornetz scalar confinement.
	The $\mathbf{S}_{t}.\mathbf{S}_{\bar{t}}$, $\mathbf{L}.\mathbf{S}$, and tensor operators $T$ in $ \left\vert J,L,S\right\rangle $ basis are given by
	\begin{eqnarray}
		T& =\Bigg \{
		\begin{array}{c}
			-\frac{1}{6(2L+3)},J=L+1 \\
			+\frac{1}{6},J=L \\
			-\frac{L+1}{6(2L-1)},J=L-1.
		\end{array}%
	\end{eqnarray}
	\noindent String tension($b$) is taken to be 0.18 $\textrm{GeV}^2$, as considered in Ref. \cite{2025}. Parameters $\alpha _{s}, \sigma$, and $m_{t}$ are taken equal to 0.1596, 0.41 $\textrm{GeV}$, and 172.42 $\textrm{GeV}$ respectively. To calculate the wave functions and mass spectrum of $t\overline{t}$, radial Schr$\ddot{\textrm{o}}$dinger is considered as
	\begin{equation}
		\frac{d^2U(r)}{dr^2}+2\mu (E-V(r))U(r)-\frac{L(L+1)}{2\mu r^{2}})U(r)=0, \label{de}
	\end{equation}
	
	where $\mu$ is the reduce mass of toponium. To find the Non-triavial solution and discrete energy state, SE is solved by shooting method.
	\begin{figure}[H]
		\centering
		{%
			\includegraphics[width=0.4\textwidth]{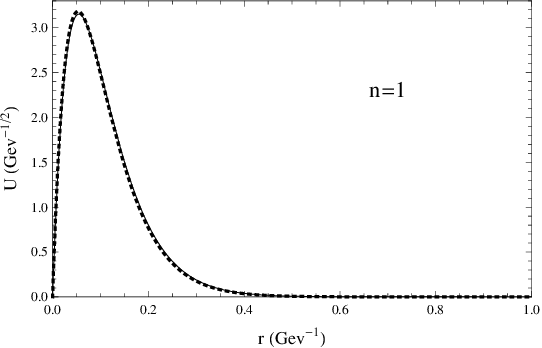} 
		}
		\hspace{0.1\textwidth} 
		{%
			\includegraphics[width=0.4\textwidth]{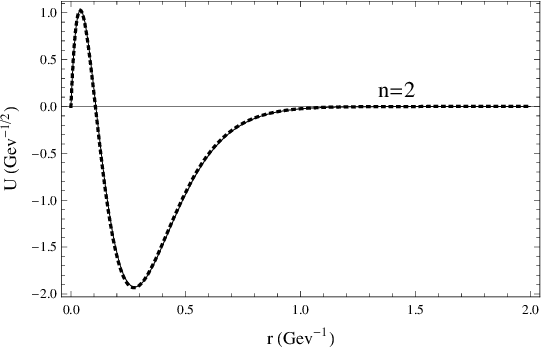} 
		}
		\hspace{0.1\textwidth} 
		{%
			\includegraphics[width=0.4\textwidth]{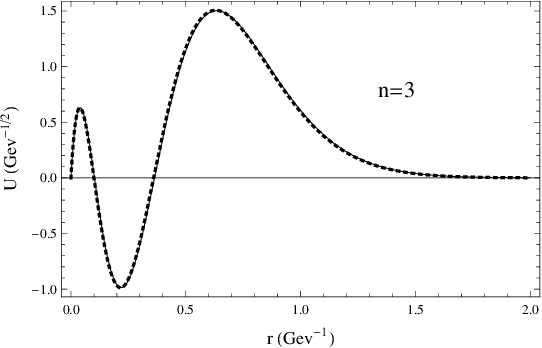} 
		}
		\hspace{0.1\textwidth} 
		{%
			\includegraphics[width=0.4\textwidth]{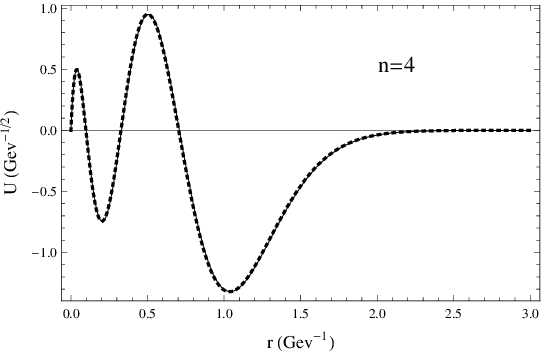} 
		}
		\caption{S-wave toponium wave functions. $\Upsilon_t$ is represented with solid curve and $\eta_t$ is represented with the dotted curve.}
		\label{fig:Swave}
	\end{figure}
	\begin{figure}[H]
		\centering
		{%
			\includegraphics[width=0.4\textwidth]{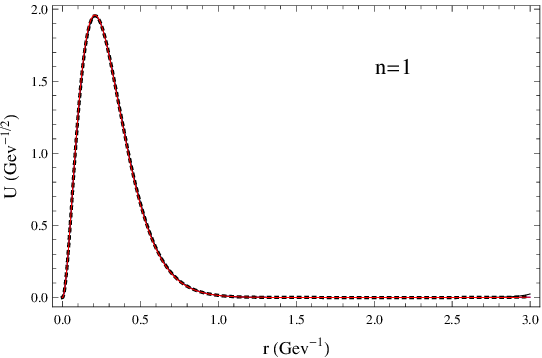} 
		}
		\hspace{0.1\textwidth} 
		{%
			\includegraphics[width=0.4\textwidth]{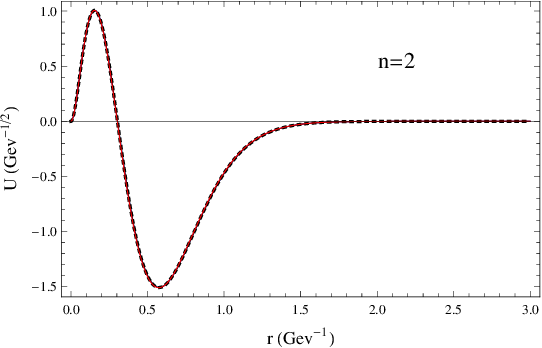} 
		}
		\hspace{0.1\textwidth} 
		{%
			\includegraphics[width=0.4\textwidth]{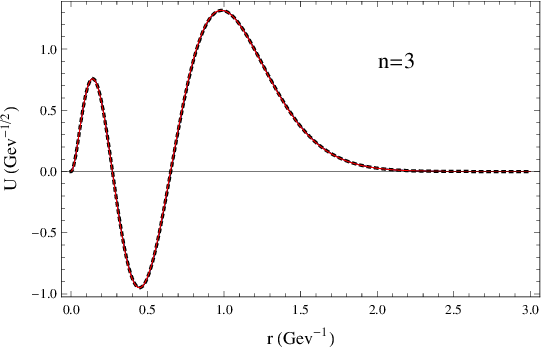} 
		}
		\hspace{0.1\textwidth} 
		{%
			\includegraphics[width=0.4\textwidth]{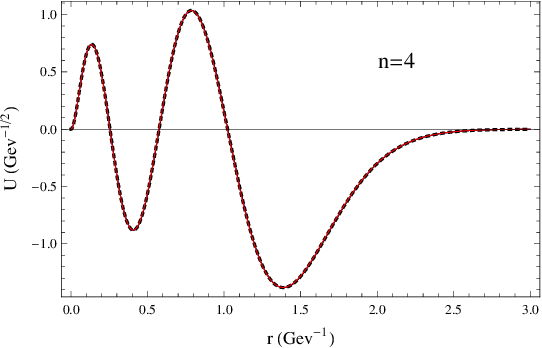} 
		}
		\caption{P-wave toponium wave functions. $\chi_{t2}$, $\chi_{t1}$, $\chi_{t0}$ and $h_{t}$ are represented with black solid line, black dotted line,red solid line and blue solid line curves respectively.}
		\label{fig:Pwave}
	\end{figure}
	\begin{figure}[H]
		\centering
		{%
			\includegraphics[width=0.4\textwidth]{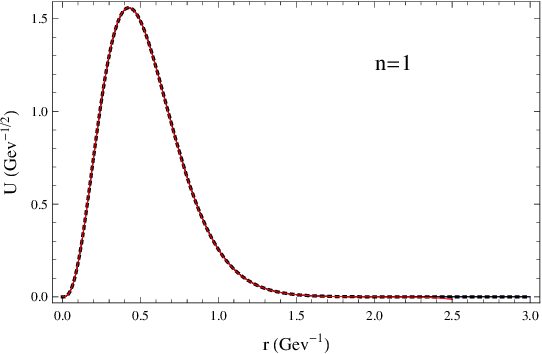} 
		}
		\hspace{0.1\textwidth} 
		{%
			\includegraphics[width=0.4\textwidth]{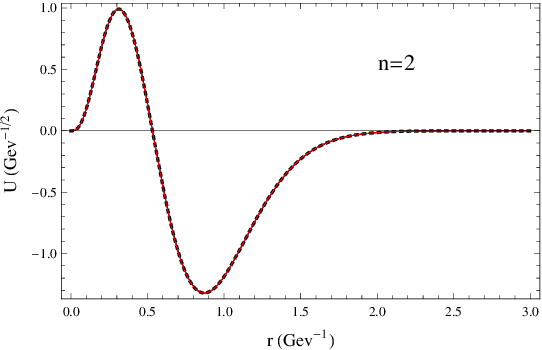} 
		}
		\hspace{0.1\textwidth} 
		{%
			\includegraphics[width=0.4\textwidth]{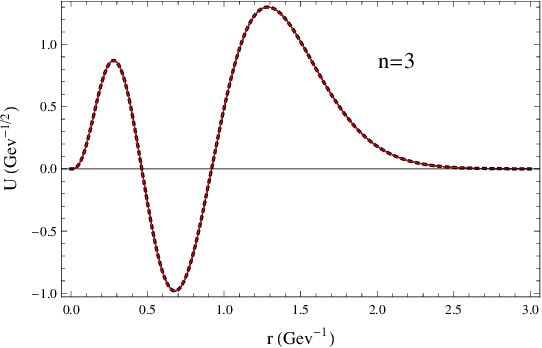} 
		}
		\hspace{0.1\textwidth} 
		{%
			\includegraphics[width=0.4\textwidth]{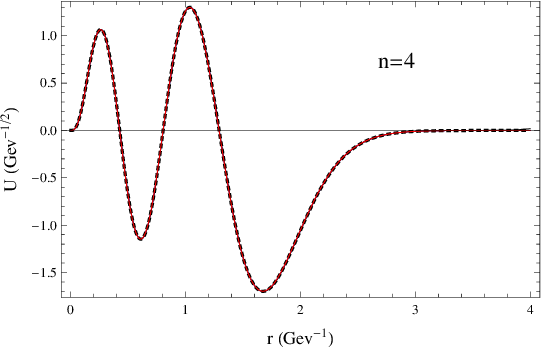} 
		}
		\caption{D-wave toponium wave functions. $\Upsilon_{t3}$, $\Upsilon_{t2}$, $\Upsilon_{t1}$ and $\eta_{t2}$ are represented with solid black line, dotted black line, solid red line and solid blue line curves respectively.}
		
		\label{fig:Dwave}
	\end{figure}
	
	Mass of different states of mesons is calculated by using following expression:
	\begin{equation}
		m_{t\bar{t}} = 2 m_{t} + E,
	\end{equation}
	
	\begin{table}[H]
		\centering
		\renewcommand{\arraystretch}{0.6}
		\caption{Masses of ground and excited states of toponium mesons.} \label{topoonium masses}
		\begin{tabular}{c c c c c c c c}
			\hline\hline
			\hspace{0.3cm} nL \hspace{0.3cm} & \hspace{0.3cm} Meson \hspace{0.3cm} & \hspace{0.2cm} Our calculated mass \hspace{0.2cm} & \hspace{0.3cm} Theo. mass~\cite{2025}  & \hspace{0.3cm} Theo. mass~\cite{2024a} & Radii \\
			\hspace{0.3cm}  \hspace{0.3cm} & \hspace{0.3cm}  \hspace{0.3cm} & \hspace{0.2cm} (\textrm{GeV}) \hspace{0.2cm} & \hspace{0.3cm} (\textrm{GeV}) & \hspace{0.3cm} (\textrm{GeV}) &$\textrm{fm}$ \\ \hline
			\multirow{2}[0]{*}{1S} & $\eta_t(1^1S_0)$    &342.867   & 343.62 & 341.267 & 0.01823\\
			& $\Upsilon_t(1^3S_1)$  &342.914   & - & 341.650 & 0.01861 \\
			\hline
			\multirow{2}[0]{*}{2S} & $\eta_t(2^1S_0)$    &344.405   & 344.59 & 344.179 & 0.0658 \\
			& $\Upsilon_t(2^3S_1)$  &344.411   & - & 344.227 & 0.0664\\
			\hline
			\multirow{2}[0]{*}{3S} & $\eta_t(3^1S_0)$    &344.742  &  344.93 & \multirow{2}[0]{*}{344.759} & 0.1307\\
			& $\Upsilon_t(3^3S_1)$  &344.745  & - & & 0.1314 \\
			\hline
			\multirow{2}[0]{*}{4S} & $\eta_t(4^1S_0)$    &344.91  &  - & \multirow{2}[0]{*}{345.00} & 0.1972\\
			& $\Upsilon_t(4^3S_1)$  &344.911   & - &  & 0.1979\\
			\hline
			\multirow{2}[0]{*}{5S} & $\eta_t(5^1S_0)$    &345.026  & - &\multirow{2}[0]{*}{345.153} & 0.2592 \\
			& $\Upsilon_t(5^3S_1)$  &345.027  & - & & 0.2598 \\
			\hline
			\multirow{2}[0]{*}{6S} & $\eta_t(6^1S_0)$    &345.12  & - & \multirow{2}[0]{*}{345.269} & 0.3163 \\
			& $\Upsilon_t(6^3S_1)$  &345.121  & - & & 0.3169 \\
			\hline
			
			\multirow{4}[0]{*}{1P} & $h_t(1^1P_1)$  &344.4  & \multirow{4}[0]{*}{344.39} & 344.206 & 0.0556\\
			& $\chi_{t0}(1^3P_0)$ &344.396  & & 344.181 & 0.0553 \\
			& $\chi_{t1}(1^3P_1)$ &344.399   & & 344.200 & 0.0556\\
			& $\chi_{t2}(1^3P_2)$ &344.401   & & 344.215 & 0.0558 \\
			\hline
			\multirow{4}[0]{*}{2P} & $h_t(2^1P_1)$ &344.734  & \multirow{4}[0]{*}{344.83} & 344.733 & 0.1212\\
			& $\chi_{t0}(2^3P_0)$ &344.733  & & 344.742 & 0.1208\\
			& $\chi_{t1}(2^3P_1)$ &344.734   & & 344.749 & 0.1211\\
			& $\chi_{t2}(2^3P_2)$ &344.735  & & 344.754 & 0.1214\\
			\hline
			\multirow{4}[0]{*}{3P} & $h_t(3^1P_1)$ &344.901  & \multirow{4}[0]{*}{345.06} & 344.992 & 0.1880\\
			& $\chi_{t0}(3^3P_0)$ &344.901  & & 344.987 & 0.1877\\
			& $\chi_{t1}(3^3P_1)$ &344.901  & & 344.991 & 0.1879\\
			& $\chi_{t2}(3^3P_2)$ &344.902 & & 344.994 & 0.1881\\
			\hline
			\multirow{4}[0]{*}{4P} & $h_t(4^1P_1)$ &345.018  &  \multirow{4}[0]{*}{-} &  \multirow{4}[0]{*}{-} & 0.2502\\
			& $\chi_{t0}(4^3P_0)$ &345.017  &  & & 0.2499 \\
			& $\chi_{t1}(4^3P_1)$ &345.018  &  & & 0.2501 \\
			& $\chi_{t2}(4^3P_2)$ &345.018  &  & & 0.2503 \\
			\hline
			\multirow{4}[0]{*}{5P} & $h_t(5^1P_1)$ &345.112  & \multirow{4}[0]{*}{-} &  \multirow{4}[0]{*}{-} & 0.3074\\
			& $\chi_{t0}(5^3P_0)$ &345.111  & & & 0.3072\\
			& $\chi_{t1}(5^3P_1)$ &345.112  & & & 0.3074\\
			& $\chi_{t2}(5^3P_2)$ &345.112  & & & 0.3076\\
			\hline
			\multirow{4}[0]{*}{6P} & $h_t(6^1P_1)$ &345.193  & \multirow{4}[0]{*}{-} &  \multirow{4}[0]{*}{-} & 0.3606\\
			& $\chi_{t0}(6^3P_0)$ &345.192  &  & & 0.3604\\
			& $\chi_{t1}(6^3P_1)$ &345.193  & & & 0.3606 \\
			& $\chi_{t2}(6^3P_2)$ &345.193  & & & 0.3607\\
			\hline
		\end{tabular}
	\end{table}
	
	\begin{table}[H]
		\centering
		\renewcommand{\arraystretch}{0.6}
		\caption{Masses of ground and excited states of toponium mesons.} \label{topoonium masses}
		\begin{tabular}{c c c c c c c c}
			\hline\hline
			\hspace{0.3cm} nL \hspace{0.3cm} & \hspace{0.3cm} Meson \hspace{0.3cm} & \hspace{0.2cm} Our calculated mass \hspace{0.2cm} & \hspace{0.3cm} Theo. mass~\cite{2025} & \hspace{0.3cm} Theo. mass~\cite{2024a} & Radii\\
			\hspace{0.3cm}  \hspace{0.3cm} & \hspace{0.3cm}  \hspace{0.3cm} & \hspace{0.2cm} (\textrm{GeV}) & \hspace{0.3cm} (\textrm{GeV}) & \hspace{0.3cm} (\textrm{GeV}) & $\textrm{fm}$\\ \hline
			
			\multirow{4}[0]{*}{1D} & $\eta_{t2}(1^1D_2)$ & 344.717  & \multirow{4}[0]{*}{344.72} &  344.735 & 0.1008\\
			& $\Upsilon_{t1}(1^3D_1)$& 344.716  & & 344.733 & 0.1006\\
			& $\Upsilon_{t2}(1^3D_2)$&344.717  & & 344.735 & 0.1007\\
			& $\Upsilon_{t3}(1^3D_3)$&344.717  &  & 344.736 & 0.1009\\
			\hline
			\multirow{4}[0]{*}{2D} & $\eta_{t2}(2^1D_2)$ &344.884  & \multirow{4}[0]{*}{344.99} & 344.976 & 0.17\\
			& $\Upsilon_{t1}(2^3D_1)$& 344.884  & & 344.975 & 0.1697\\
			& $\Upsilon_{t2}(2^3D_2)$& 344.884  & & 344.976 & 0.1699\\
			& $\Upsilon_{t3}(2^3D_3)$&344.884  & & 344.977 & 0.1701 \\
			\hline
			\multirow{4}[0]{*}{3D} & $\eta_{t2}(3^1D_2)$ &345.002  & \multirow{4}[0]{*}{345.16} & 345.130 & 0.2335 \\
			& $\Upsilon_{t1}(3^3D_1)$&345.001  & & 345.129 & 0.2332\\
			& $\Upsilon_{t2}(3^3D_2)$& 345.001  & & 345.130 & 0.2334\\
			& $\Upsilon_{t3}(3^3D_3)$&345.002  & & 345.131 & 0.2336\\
			\hline
			\multirow{4}[0]{*}{4D} & $\eta_{t2}(4^1D_2)$ &345.096 & \multirow{4}[0]{*}{-} & \multirow{4}[0]{*}{-} & 0.2916\\
			& $\Upsilon_{t1}(4^3D_1)$&345.096  & & & 0.2914\\
			& $\Upsilon_{t2}(4^3D_2)$& 345.096  & & & 0.2903\\
			& $\Upsilon_{t3}(4^3D_3)$&345.096  & & & 0.2917\\
			\hline
			
			\multirow{4}[0]{*}{5D} & $\eta_{t2}(4^1D_2)$ &345.177  & \multirow{4}[0]{*}{-} & \multirow{4}[0]{*}{-}& 0.3454\\
			& $\Upsilon_{t1}(5^3D_1)$&345.177  & & & 0.3452\\
			& $\Upsilon_{t2}(5^3D_2)$& 345.177  & & & 0.3454\\
			& $\Upsilon_{t3}(5^3D_3)$&345.177  & & &0.3456\\
			\hline
			\
			\multirow{4}[0]{*}{1F} & $h_{t3}(1^1F_3)$  &344.93 &  \multirow{4}[0]{*}{344.8316} & \multirow{4}[0]{*}{-} & 0.1460\\
			& $\chi_{t2}(1^3F_2)$ &344.862  & & & 0.1442\\
			& $\chi_{t3}(1^3F_3)$ &344.862  & & &0.1445 \\
			& $\chi_{t4}(1^3F_4)$ &344.862   & & & 0.1462\\
			\hline
			\multirow{4}[0]{*}{2F} & $h_{t3}(2^1F_3)$ &344.98  & \multirow{4}[0]{*}{345.12} & \multirow{4}[0]{*}{-} & 0.2112\\
			& $\chi_{t2}(2^3F_2)$ & 344.98 & & & 0.2111 \\
			& $\chi_{t3}(2^3F_3)$ &344.98  & & & 0.2112\\
			& $\chi_{t4}(2^3F_4)$ &344.981  & & & 0.2113\\
			\hline
			\multirow{4}[0]{*}{3F} & $h_{t3}(3^1F_3)$ & 345.076  & \multirow{4}[0]{*}{345.25} & \multirow{4}[0]{*}{-} & 0.2714\\
			& $\chi_{t2}(3^3F_2)$ &345.076 & & & 0.2715\\
			& $\chi_{t3}(3^3F_3)$ & 345.076 & & & 0.2727\\
			& $\chi_{t4}(3^3F_4)$ &345.076 & & & 0.2714\\
			\hline
		\end{tabular}
	\end{table}
	\section{Radiative transitions}
	Radiative transitions play an important role in the investigation of the higher states of $t\overline{t}$ mesons. $E1$ radiative transitions from one $t\overline{t}$ state to other state are found by using the following expression which is given in ref.\cite{Godfrey05} as:
	\begin{equation}
		\Gamma_{E1}(n^{2S+1}L_J\rightarrow n'^{2S'+1}L'_{J'}+\gamma)=\frac{4}{3}C_{fi}\delta_{S S'}e_t^2 \alpha\mid < \Psi_f \mid r \mid \Psi_i>\mid^2 E_\gamma^3 \frac{E^{(t\overline{t})}_f}{M^{(t \overline{t})}_i}.  \label{E1}
	\end{equation}
	Here
	$E_\gamma= \frac{M_i^2 - M_f^2}{2 M_i}$ is the final photon energy,
	$E^{t \overline{t}}_f$ is the energy of the final toponium meson state, $M_i$ is the initial state mass of $t\overline{t}$ meson, and $C_{fi}$ matrix element expression is
	\begin{equation}
		C_{fi}=max(L, L')(2 J'+1)\left \{
		\begin{array}{ccc}
			L' & J' & S \\
			J & L & 1 \\
		\end{array}
		\right \}^2.
	\end{equation}
	$M1$ radiative transitions from one $t\overline{t}$ state to other state with same orbital quantum number $l$ are found by using the following expression which is given in ref.\cite{Godfrey05}
	\begin{equation}
		\Gamma_{M1}(n^{2S+1}L_J\rightarrow n'^{2S'+1}L'_{J'}+\gamma)=\frac{4}{3}\frac{2J'+1}{2 L+1}\delta_{L L'}\delta_{S S'\pm 1}e_t^2 \frac{\alpha}{m^2_t}\mid < \Psi_f \mid \Psi_i>\mid^2 E_\gamma^3\frac{E^{(t \overline{t})}_f}{M^{(t \overline{t})}_i}. \label{M1}
	\end{equation}
	\begin{table}[h!]
		\centering
		\begin{tabular}{c c c c c c}
			\hline
			\hline
			Initial State&Final State&$E_\gamma $ (MeV)&Predicted Width(KeV)&Predicted B.R($\%$)\\
			\hline\hline
			$\Upsilon_t (1 ^3S_1) $ & $\eta_t(1 ^1S_0)\gamma$  &46.997 &$1.51\times 10^{-5}$  & 100  \\
			\hline
			$\eta_t (2 ^1S_0) $         & $h_t(1 ^1P_1)\gamma$ & 4.999 &0.00470 &0.35   \\
			&  $\Upsilon_t(1 ^3S_1)\gamma$   &  1487.77&1.33895&99.64  \\
			&Total& &1.34366&100 \\
			\hline
			$\Upsilon_t (2 ^3S_1) $       &  $\chi_{t2}(1 ^3P_2)\gamma$   &9.999 &$0.02071$& 3.80  \\
			&  $\chi_{t1}(1 ^3P_1)\gamma$   & 11.999&$0.02147$& 3.94 \\
			&  $\chi_{t0}(1 ^3P_0)\gamma$   & 14.999&$0.01398$ & 2.56\\
			&  $\eta_{t}(2 ^1S_0)\gamma$   &5.999&$3.13\times10^{-8}$&$5.75\times 10^{-6} $     \\
			&  $\eta_{t}(1 ^1S_0)\gamma$   & 1540.54&0.48861&89.6 \\
			&Total& &0.54477 &100\\
			\hline
			$\eta_t (3 ^1S_0) $         &  $h_t(2 ^1P_1)\gamma$   & 7.999&0.00038&99.65   \\
			&  $h_t(1 ^1P_1)\gamma$   & 341.83&$1.27\times10^{-6}$& 0.33  \\
			&  $\Upsilon_t(2 ^3S_1)\gamma$   &330.841&$3.22\times10^{-8}$& 0.084   \\
			&  $\Upsilon_t(1 ^3S_1)\gamma$   & 1823.15&$1.96\times10^{-9}$& $5.14\times10^{-4}$  \\
			&Total& &0.00038&100\\
			\hline
			$\Upsilon_t (3 ^3S_1) $       &  $\chi_{t2}(2 ^3P_2)\gamma$   &9.999&$0.00025$&47.58   \\
			&  $\chi_{t1}(2 ^3P_1)\gamma$   &10.999&0.00019 & 36.16 \\
			&  $\chi_{t0}(2 ^3P_0)\gamma$   & 11.9998&$8.47\times 10^{-5}$& 16.12 \\
			&  $\chi_{t2}(1 ^3P_2)\gamma$   & 343.828&$1.18\times10^{-13}$&$2.24\times10^{-8}$
			\\
			&  $\chi_{t1}(1 ^3P_1)\gamma$   & 345.826&$7.23\times10^{-14}$& $1.37\times 10^{-8}$
			\\
			&  $\chi_{t0}(1 ^3P_0)\gamma$   &  348.823&$2.47\times10^{-14}$&$4.7\times  10^{-9}$ \\
			&  $\eta_{t}(3 ^1S_0)\gamma$   &2.999&$3.93\times10^{-9}$& 0.00074  \\
			&  $\eta_{t}(2 ^1S_0)\gamma$   & 339.832 &$1.79\times10^{-8}$&0.000034  \\
			&  $\eta_{t}(1 ^1S_0)\gamma$   & 1872.88&$7.27\times10^{-7}$& 0.1383   \\
			&Total&  &0.00052&100\\
			\hline
		\end{tabular}
		\caption{Radiative widths and branching ratios for 1S, 2S and 3S toponium mesons.}\label{results-1}
	\end{table}
	\begin{table}[h!]
		\centering
		\begin{tabular}{c c c c c c}
			\hline
			\hline
			Initial State&Final State&$E_\gamma $ (MeV)&Predicted Width(KeV)&Predicted B.R($\%$)\\
			\hline\hline
			$\eta_t (4 ^1S_0) $         &  $h_t(3 ^1P_1)\gamma$   & 8.999&0.00249 &  1.95\\
			&  $h_t(2 ^1P_1)\gamma$   &175.955&$0.12477$ & 98.03 \\
			&  $h_t(1 ^1P_1)\gamma$   &  509.623  &$7.68\times10^{-6}$&0.0060\\
			&  $\Upsilon_t(3 ^3S_1)\gamma$   &164.961&$2.50\times10^{-7}$&0.00019   \\
			&  $\Upsilon_t(2 ^3S_1)\gamma$  &498.639&$2.79\times10^{-8}$ &0.000021   \\
			&  $\Upsilon_t(1 ^3S_1)\gamma$   &1990.22&$1.85\times10^{-9}$&$1.46\times10^{-6}$ \\
			&Total& &0.12726&100\\
			\hline
			$\Upsilon_t (4 ^3S_1) $       &  $\chi_{t2}(3 ^3P_2)\gamma$   &8.999&0.00139& 0.99  \\
			&  $\chi_{t1}(3 ^3P_1)\gamma$   & 9.999& 0.00114& 0.81 \\
			&  $\chi_{t0}(3 ^3P_0)\gamma$   & 9.999& 0.00038&0.27\\
			&  $\chi_{t2}(2 ^3P_2)\gamma$   & 175.955& 0.04784&34.29 \\
			&  $\chi_{t1}(2 ^3P_1)\gamma$   & 176.955& 0.02919&20.92 \\
			&  $\chi_{t0}(2 ^3P_0)\gamma$   &  177.954&0.00990& 7.09 \\
			&  $\chi_{t2}(1 ^3P_2)\gamma$   & 509.623&$$0.02738& 19.63\\
			&  $\chi_{t1}(1 ^3P_1)\gamma$   & 511.62&$$0.01662&11.91\\
			&  $\chi_{t0}(1 ^3P_0)\gamma$   &  514.616&$0.00564$& 4.04\\
			&  $\eta_{t}(4 ^1S_0)\gamma$   &0.999&$1.46\times10^{-10}$&$1.04\times 10^{-7}$  \\
			&  $\eta_{t}(3 ^1S_0)\gamma$   &168.959&$8.60\times10^{-8}$&$0.000061$  \\
			&  $\eta_{t}(2 ^1S_0)\gamma$   & 505.629&$1.36\times10^{-7}$& 0.000097\\
			&  $\eta_{t}(1 ^1S_0)\gamma$   &2037.94&$1.03\times 10^{-5}$ & 0.00738\\
			&Total&  &0.13949&100\\
			\hline
		\end{tabular}
		\caption{Radiative widths and branching ratios for 4S toponium mesons.}\label{results-2}
	\end{table}
	\begin{table}[h!]
		\centering
		\begin{tabular}{c c c c c c}
			\hline
			\hline
			Initial State&Final State&$E_\gamma $ (MeV)&Predicted Width(KeV)&Predicted B.R($\%$)\\
			\hline\hline
			$\eta_t (5 ^1S_0) $        &  $h_t(4 ^1P_1)\gamma$   & 7.999&0.00307&1.167   \\
			&  $h_t(3 ^1P_1)\gamma$   &124.977&0.192717 &73.26   \\
			&  $h_t(2 ^1P_1)\gamma$   &  291.876&0.06012& 22.85 \\
			&  $h_t(1 ^1P_1)\gamma$   &  625.432&0.00716& 2.72\\
			&  $\Upsilon_t(4 ^3S_1)\gamma$   & 114.981&$8.23\times10^{-8}$& 0.
			000031  \\
			&  $\Upsilon_t(3 ^3S_1)\gamma$   &280.886&$3.22\times10^{-7}$ & 0.00012  \\
			&  $\Upsilon_t(2 ^3S_1)\gamma$   & 614.452&$2.16\times10^{-7}$& 0
			.000082 \\
			&  $\Upsilon_t(1 ^3S_1)\gamma$   & 2105.54&$9.71\times10^{-7}$&0.000369  \\
			&Total &  &0.2630&100\\
			\hline
			$\Upsilon_t (5 ^3S_1) $       &  $\chi_{t2}(4 ^3P_2)\gamma$  & 8.999 &0.00241&  0.99  \\
			&  $\chi_{t1}(4 ^3P_1)\gamma$     &8.999&0.00144& 0.59\\
			&  $\chi_{t0}(4 ^3P_0)\gamma$     &9.999    &0.00066&0.27 \\
			&  $\chi_{t2}(3 ^3P_2)\gamma$      & 124.977  &0.09560&39.35\\
			&  $\chi_{t1}(3 ^3P_1)\gamma$     &125.977  &0.05875&24.18\\
			&  $\chi_{t0}(3 ^3P_0)\gamma$     &125.977  &0.01958 & 8.06 \\
			&  $\chi_{t2}(2 ^3P_2)\gamma$     &291.876&0.03137 &12.91\\
			&  $\chi_{t1}(2 ^3P_1)\gamma$    &292.876&0.01902 & 7.83\\
			&  $\chi_{t0}(2 ^3P_0)\gamma$    &621.439&0.00640 &2.63\\
			&  $\chi_{t2}(1 ^3P_2)\gamma$     &625.432&0.00424& 1.74\\
			&  $\chi_{t1}(1 ^3P_1)\gamma$     &627.428&0.00257& 1.05\\
			&  $\chi_{t0}(1 ^3P_0)\gamma$     &630.423   &0.00087& 0.35\\
			&  $\eta_{t}(5 ^1S_0)\gamma$   &0.999&$1.45\times10^{-10}$&$5.96\times 10^{-8}$   \\
			&  $\eta_{t}(4 ^1S_0)\gamma$   &116.98&$2.77\times 10^{-8}$& 0.000011  \\
			&  $\eta_{t}(3 ^1S_0)\gamma$   &284.882&$1.10\times10^{-7}$& 0.000045  \\
			&  $\eta_{t}(2 ^1S_0)\gamma$   &621.439&$1.41\times10^{-8}$ & $5.86\times 10^{-6}$ \\
			&  $\eta_{t}(1 ^1S_0)\gamma$   &2153.24&$3.62\times10^{-7}$ &0.00015  \\
			&Total & &0.2429&100\\
			\hline
		\end{tabular}
		\caption{Radiative widths and branching ratios for 5S toponium mesons.}\label{results-3}
	\end{table}
	\begin{table}[h!]
		\centering
		\begin{tabular}{c c c c c c}
			\hline
			\hline
			Initial State&Final State&$E_\gamma $ (MeV)&Predicted Width(KeV)&Predicted B.R($\%$)\\
			\hline\hline
			$\eta_t (6 ^1S_0) $        &$h_t(5 ^1P_1)\gamma$   &7.999&0.00457 &0.192 \\
			&  $h_t(4 ^1P_1)\gamma$   & 101.985&0.17291&7.299 \\
			&  $h_t(3 ^1P_1)\gamma$   &218.931&0.10154 & 4.286  \\
			&  $h_t(2 ^1P_1)\gamma$   & 385.784&0.50754 &21.42\\
			&  $h_t(1 ^1P_1)\gamma$   & 719.249&1.58223&66.79\\
			&  $\Upsilon_t(5 ^3S_1)\gamma$   &92.987&$4.28\times10^{-8}$ &$1.80\times10^{-6}$   \\
			&  $\Upsilon_t(4 ^3S_1)\gamma$   & 208.937&$1.29\times10^{-7}$& $5.44\times10^{-6}$  \\
			&  $\Upsilon_t(3 ^3S_1)\gamma$   &  374.796&$3.58\times10^{-7}$& $0.000015$ \\
			&  $\Upsilon_t(2 ^3S_1)\gamma$   &  708.272&$1.73\times 10^{-5}$&0.000730\\
			&  $\Upsilon_t(1 ^3S_1)\gamma$   & 2198.95 &0.00019&0.00802\\
			&Total &  &2.36881&100\\
			\hline
			$\Upsilon_t (6 ^3S_1) $       &  $\chi_{t2}(5 ^3P_2)\gamma$  &8.999&0.003597& 0.16 \\
			&  $\chi_{t1}(5 ^3P_1)\gamma$     &8.999&0.002158&0.09\\
			&  $\chi_{t0}(5 ^3P_0)\gamma$     &9.999&0.0009868&0.04 \\
			&  $\chi_{t1}(4 ^3P_1)\gamma$     &102.985&0.053157&2.37\\
			&  $\chi_{t0}(4 ^3P_0)\gamma$     & 103.984&0.01824&0.81\\
			&  $\chi_{t2}(3 ^3P_2)\gamma$      &218.931&0.060088&2.71 \\
			&  $\chi_{t1}(3 ^3P_1)\gamma$     &219.93&0.036548&1.63\\
			&  $\chi_{t0}(3 ^3P_0)\gamma$     & 219.93&0.012182& 0.54\\
			&  $\chi_{t2}(2 ^3P_2)\gamma$     &385.784&0.094768&4.23 \\
			&  $\chi_{t1}(2 ^3P_1)\gamma$    & 719.249&0.368127&16.43 \\
			&  $\chi_{t0}(2 ^3P_0)\gamma$    & 387.782&0.019249&0.85\\
			&  $\chi_{t2}(1 ^3P_2)\gamma$     & 719.249&0.92065&41.10\\
			&  $\chi_{t1}(1 ^3P_1)\gamma$     & 721.245&0.557003&54.86\\
			&  $\chi_{t0}(1 ^3P_0)\gamma$     &724.238&0.187987 &8.39\\
			&  $\eta_{t}(6 ^1S_0)\gamma$   &  0.999&$1.45\times10^{-10}$&$6.51\times 10^{-7}$ \\
			&  $\eta_{t}(5 ^1S_0)\gamma$   &   94.987&$1.46\times10^{-8}$&$5.44\times 10^{-6}$\\
			&  $\eta_{t}(4 ^1S_0)\gamma$   &  210.935&$4.23\times10^{-8}$&$1.88\times 10^{-6}$\\
			&  $\eta_{t}(3 ^1S_0)\gamma$   &   378.792&$1.23\times10^{-7}$&$5.4\times 10^{-7}$\\
			&  $\eta_{t}(2 ^1S_0)\gamma$   &   715.257&$7.0452\times10^{-6}$&0.00031\\
			&  $\eta_{t}(1 ^1S_0)\gamma$   &   2246.64&$0.000072$&0.0032\\
			&Total&&2.2400&100\\
			
			\hline
			
		\end{tabular}
		\caption{Radiative widths and branching ratios for 6S toponium mesons.}\label{results-3}
	\end{table}
	\begin{table}[h!]
		\centering
		\begin{tabular}{c c c c c c}
			\hline
			\hline
			Initial State&Final State&$E_\gamma $ (MeV)&Predicted Width(KeV)&Predicted B.R($\%$)\\
			\hline\hline
			$h_t (1 ^1P_1) $            &  $\chi_{t0}(1 ^3P_0)\gamma$   &3.999&$3.10\times10^{-9}$&  $1.224\times 10^{-8}$  \\
			&  $\chi_{t1}(1 ^3P_1)\gamma$   &0.999&$1.45\times10^{-10}$&   $5.72 \times 10^{-10}$  \\
			&  $\eta_t(1 ^1S_0)\gamma$   & 1529.59&25.3268 &  100 \\
			&Total& &25.32&100\\
			\hline
			$\chi_{t0} (1 ^3P_0) $      &  $\Upsilon_t(1 ^3S_1)\gamma$   &1478.81&24.6204    &100  \\
			\hline
			$\chi_{t1} (1 ^3P_1) $      &  $\Upsilon_t(1 ^3S_1)\gamma$   &1481.8&24.7697  &100  \\
			\hline
			$\chi_{t2} (1 ^3P_2) $      &  $h_t(1 ^1P_1)\gamma$   &  0.999&$1.45\times10^{-10}$&$5.27\times 10^{-10}$ \\
			&  $\Upsilon_t(1 ^3S_1)\gamma$   &1483.79&24.8695 & 100 \\
			&Total&  &25.3268&100\\
			\hline
		\end{tabular}
		\caption{Radiative widths and branching ratios for 1P toponium mesons.}\label{results-4}
	\end{table}
	\begin{table}[h!]
		\centering
		\begin{tabular}{c c c c c c}
			\hline
			\hline
			Initial State&Final State&$E_\gamma $ (MeV)&Predicted Width(KeV)&Predicted B.R($\%$)\\
			\hline\hline
			$h_t (2 ^1P_1) $            &  $\eta_{t2}(1 ^1D_2)\gamma$   & 16.999&0.00271&  0.178 \\
			&  $\chi_{t0}(2 ^3P_0)\gamma$   &0.999&$1.84\times10^{-12}$&$1.20\times 10^{-10}$     \\
			&  $\chi_{t0}(1 ^3P_0)\gamma$   & 337.834&0.00090 & 0.05916\\
			&  $\chi_{t1}(1 ^3P_1)\gamma$   & 334.837&0.00264 &  0.1735  \\
			&  $\chi_{t2}(1 ^3P_2)\gamma$   & 334.837&0.00440 &   0.2899 \\
			&  $\eta_t(2 ^1S_0)\gamma$   &  328.843&1.46125  &96.05\\
			&  $\eta_t(1 ^1S_0)\gamma$   &  328.843&0.04939&3.25\\
			&Total&  &1.5212&100\\
			\hline
			$\chi_{t0} (2 ^3P_0) $      &  $\Upsilon_{t1}(1 ^3D_1)\gamma$   &16.999&0.00271  &0.0364  \\
			&  $h_t(1 ^1P_1)\gamma$   &  332.839&0.00337  & 0.0453\\
			&  $\Upsilon_t(2 ^3S_1)\gamma$   & 321.85&1.4652 & 19.71  \\
			&  $\Upsilon_t(1 ^3S_1)\gamma$   & 1814.2&5.96174& 80.20\\
			&Total& &7.43&100\\
			\hline
			$\chi_{t1} (2 ^3P_1) $      &  $\Upsilon_{t2}(1 ^3D_2)\gamma$   & 16.999&0.00203  &0.02722   \\
			&  $\Upsilon_{t1}(1 ^3D_1)\gamma$   &17.999&0.000805  & 0.01079 \\
			&  $h_t(1 ^1P_1)\gamma$   &   333.838&0.00372  & 0.04989\\
			&  $\Upsilon_t(2 ^3S_1)\gamma$   &322.849&1.47888 &19.8322\\
			&  $\Upsilon_t(1 ^3S_1)\gamma$   & 1815.2&5.97154 &  80.0799 \\
			&Total& &7.4569&100\\
			\hline
			$\chi_{t2} (2 ^3P_2) $      &  $\Upsilon_{t3}(1 ^3D_3)\gamma$   & 17.999&0.00270& 0.03607 \\
			&  $\Upsilon_{t2}(1 ^3D_2)\gamma$   & 17.999&0.00048& 0.00641 \\
			&  $\Upsilon_{t1}(1 ^3D_1)\gamma$   & 18.999&0.00284&   0.03794  \\
			&  $h_t(2 ^1P_1)\gamma$   &0.999&$5.52\times10^{-12}$&$7.35\times 10^{-11}$    \\
			&  $h_t(1 ^1P_1)\gamma$   & 334.837&0.00376 &0.05202\\
			&  $\Upsilon_t(2 ^3S_1)\gamma$   & 323.848&1.49265&19.94    \\
			&  $\Upsilon_t(1 ^3S_1)\gamma$   &  1816.19&5.98134& 79.92 \\
			&Total& &7.48377&100\\
			\hline
		\end{tabular}
		\caption{Radiative widths and branching ratios for 2P toponium mesons.}\label{results-5}
	\end{table}
	\begin{table}[h!]
		\centering
		\begin{tabular}{c c c c c c}
			\hline
			\hline
			Initial State&Final State&$E_\gamma $ (MeV)&Predicted Width(KeV)&Predicted B.R($\%$)\\
			\hline\hline
			$h_t (3 ^1P_1) $            &  $\eta_{t2}(2 ^1D_2)\gamma$   &16.999&0.00680 &0.103 \\
			&  $\eta_{t2}(1 ^1D_2)\gamma$   & 16.999&0.00026 &0.00395  \\
			&  $\chi_{t0}(2 ^3P_0)\gamma$   &   167.959&$7.01\times 10^{-7}$ &0.000010  \\
			&  $\chi_{t1}(2 ^3P_1)\gamma$   &  166.96&$2.06\times10^{-6}$  &0.000031\\
			&  $\chi_{t2}(2 ^3P_2)\gamma$   &   165.96&$3.382\times 10^{-6}$& 0.000051  \\
			&  $\chi_{t0}(1 ^3P_0)\gamma$   &   504.63&$2.64\times 10^{-5}$&0.000401\\
			&  $\chi_{t1}(1 ^3P_1)\gamma$   &   501.635& $7.78\times 10^{-5}$&0.00118  \\
			&  $\chi_{t2}(1 ^3P_2)\gamma$   &    499.638&0.00013&0.00197\\
			&  $\eta_t(3 ^1S_0)\gamma$   &  158.963&0.47420 & 7.213  \\
			&  $\eta_t(2 ^1S_0)\gamma$   &  495.643&6.09231& 92.67 \\
			&  $\eta_t(1 ^1S_0)\gamma$   &   504.63&$2.64\times 10^{-5}$&0.00040  \\
			&Total& & 6.57 &100\\
			\hline
			$\chi_{t0} (3 ^3P_0) $      &  $\Upsilon_{t1}(2 ^3D_1)\gamma$   & 16.999&0.00804&0.9869\\
			&  $\Upsilon_{t1}(1 ^3D_1)\gamma$   &184.95&0.02999&0.3681\\
			&  $h_t(2 ^1P_1)\gamma$   & 166.96&$5.69\times10^{-9}$&$6.98\times 10^{-8}$  \\
			&  $h_t(1 ^1P_1)\gamma$   & 500.636&$3.77\times 10^{-7}$& $4.62\times 10^{-6}$\\
			&  $\Upsilon_t(3 ^3S_1)\gamma$   &  155.965&0.47930 &  5.88\\
			&  $\Upsilon_t(2 ^3S_1)\gamma$   &  489.652&0.87728& 10.76\\
			&  $\Upsilon_t(1 ^3S_1)\gamma$   &  1981.28&6.75143&  82.87\\
			&Total&   &8.146\\
			\hline
			$\chi_{t1} (3 ^3P_1) $      &  $\Upsilon_{t2}(2 ^3D_2)\gamma$   &16.9996&0.00603&0.07402   \\
			&  $\Upsilon_{t1}(2 ^3D_1)\gamma$   &16.9996&0.00201&0.02467 \\
			&  $\Upsilon_{t2}(1 ^3D_2)\gamma$   &183.951&0.02213&0.27167 \\
			&  $\Upsilon_{t1}(1 ^3D_1)\gamma$   &184.95&0.00750 & 0.092073 \\
			&  $h_t(2 ^1P_1)\gamma$   &166.96&$5.69\times10^{-9}$&  $6.98\times 10^{-8}$  \\
			&  $h_t(1 ^1P_1)\gamma$ & 500.636&$3.77\times 10^{-7}$&$4.62\times 10^{-6}$ \\
			&  $\Upsilon_t(3 ^3S_1)\gamma$   & 155.965&0.47929 & $5.883$\\
			&  $\Upsilon_t(2 ^3S_1)\gamma$   &489.652&0.87727  & $1.07$\\
			&  $\Upsilon_t(1 ^3S_1)\gamma$   & 1981.28&6.75143 & 82.88 \\
			&Total& &8.14&100\\
			\hline
			$\chi_{t2} (3 ^3P_2) $      &  $\Upsilon_{t3}(2 ^3D_3)\gamma$   &17.9995&0.008018 &0.09811   \\
			&  $\Upsilon_{t2}(2 ^3D_2)\gamma$   & 17.9995&0.001431& 0.01751  \\
			&  $\Upsilon_{t1}(2 ^3D_1)\gamma$   &  17.9995&0.000095& 0.00116  \\
			&  $\Upsilon_{t3}(1 ^3D_3)\gamma$   &  184.95&0.025190 & 0.3082\\
			&  $\Upsilon_{t2}(1 ^3D_2)\gamma$   &  184.95&0.004498 &0.05503\\
			&  $\Upsilon_{t1}(1 ^3D_1)\gamma$   &  185.95&0.000305 &0.00373\\
			&  $h_t(3 ^1P_1)\gamma$   &0.999&$1.44\times10^{-10}$& $1.76\times 10^{-9}$\\
			&  $h_t(2 ^1P_1)\gamma$   &167.959&$5.793\times10^{-9}$& $7.08\times 10^{-8}$ \\
			&  $h_t(1 ^1P_1)\gamma$   & 501.635&$3.80\times10^{-7}$ &$4.64\times 10^{-6}$ \\
			&  $\Upsilon_t(3 ^3S_1)\gamma$   & 156.964&0.48857& 5.978\\
			&  $\Upsilon_t(2 ^3S_1)\gamma$   & 490.651&0.88265& 10.80 \\
			&  $\Upsilon_t(1 ^3S_1)\gamma$   & 1982.27&6.76158&82.73\\
			&Total& &8.1723&100\\
			\hline
		\end{tabular}
		\caption{Radiative widths and branching ratios for 3P toponium mesons.}\label{results-6}
	\end{table}
	\begin{table}[h!]
		\centering
		\begin{tabular}{c c c c c c}
			\hline
			\hline
			Initial State&Final State&$E_\gamma $ (MeV)&Predicted Width(KeV)&Predicted B.R($\%$)\\
			\hline\hline
			$\eta_{t2} (1 ^1D_2) $      &  $h_t(1 ^1P_1)\gamma$   &483.66&8.28242 & 99.99 \\
			&  $\Upsilon_{t1}(1 ^3D_1)\gamma$   &167.959&0.00013 &0.00156  \\
			&  $\Upsilon_{t2} (1 ^3D_2)\gamma$   &  166.96&0.00021 &0.002535 \\
			&Total& &8.282&100\\
			\hline
			$\Upsilon_{t1} (1 ^3D_1) $     &  $\chi_{t0}(1 ^3P_0)\gamma$   & 319.851&2.08077& 56.28   \\
			&  $\chi_{t1}(1 ^3P_1)\gamma$   &  316.854&1.51713 &  41.03 \\
			&  $\chi_{t2}(1 ^3P_2)\gamma$   &  314.856&0.09924 & 2.684\\
			&Total&  &3.69&100\\
			\hline
			$\Upsilon_{t2} (1 ^3D_2) $     &  $\chi_{t1}(1 ^3P_1)\gamma$   &  317.853&2.75674 & 75.35  \\
			&  $\chi_{t2}(1 ^3P_2)\gamma$   &    315.855&0.90170& 24.64 \\
			&Total& &  &100\\
			\hline
			$\Upsilon_{t3} (1 ^3D_3) $     &  $\chi_{t2}(1 ^3P_2)\gamma$   &   315.855&3.60679 &100  \\
			\hline
		\end{tabular}
		\caption{Radiative widths and branching ratios for 1D toponium mesons.}\label{results-7}
	\end{table}
	\begin{table}[h!]
		\centering
		\begin{tabular}{c c c c c c}
			\hline
			\hline
			Initial State&Final State&$E_\gamma $ (MeV)&Predicted Width(KeV)&Predicted B.R($\%$)\\
			\hline\hline
			$\eta_{t2} (2 ^1D_2) $      &  $h_t(2 ^1P_1)\gamma$   & 149.967&0.94328 & 40.69   \\
			&  $h_t(1 ^1P_1)\gamma$   &483.66&1.36689  &58.96\\
			&  $\Upsilon_{t1} (1 ^3D_1)\gamma$   &167.959&$7.31\times10^{-9}$&$3.15\times 10^{-7}$     \\
			&  $\Upsilon_{t2} (1 ^3D_2)\gamma$   &166.96&$1.19\times10^{-8}$ &$5.15\times 10^{-7}$   \\
			&  $\Upsilon_{3t}(1 ^3D_3)\gamma$   & 166.96&$1.67\times10^{-8}$ & $7.20\times 10^{-7}$  \\
			&  $h_{t3} (1 ^1F_3)\gamma$   &  21.9993&0.00801 &0.3450\\
			&Total& &  &100\\
			\hline
			$\Upsilon_{t1} (2 ^3D_1) $     &  $\chi_{t0}(2 ^3P_0)\gamma$   & 150.967&0.53556& 23.157 \\
			&  $\chi_{t1}(2 ^3P_1)\gamma$   &  149.967&0.39374 &17.025\\
			&  $\chi_{t2}(2 ^3P_2)\gamma$   &  148.968&0.025728 &1.11248\\
			&  $\chi_{t0}(1 ^3P_0)\gamma$   & 487.655&0.756226&32.6991\\
			&  $\chi_{t1}(1 ^3P_1)\gamma$   &  484.659&0.556785& 24.075 \\
			&  $\chi_{t2}(1 ^3P_2)\gamma$   &   482.662&0.03666&1.5851 \\
			&  $\chi_{t2} (1 ^3F_2)\gamma$   & 21.999&0.00798   & 0.3450\\
			&Total&  &  &100\\
			\hline
			$\Upsilon_{t2} (2 ^3D_2) $  &  $\chi_{t1}(2 ^3P_1)\gamma$   & 149.967&0.70874&30.89   \\
			&  $\chi_{t2}(2 ^3P_2)\gamma$   &  148.968&0.23155 & 10.13 \\
			&  $\chi_{t1}(1 ^3P_1)\gamma$   &  484.659&1.00221 & 43.86 \\
			&  $\chi_{t2}(1 ^3P_2)\gamma$   &   482.662&0.32996& 14.44 \\
			&  $\chi_{t2} (1 ^3F_2)\gamma$   & 21.9993&0.00798&0.3493    \\
			&  $\chi_{t3} (1 ^3F_3)\gamma$   & 21.9993&0.00709 & 0.3103\\
			&Total& &2.2845  &100\\
			\hline
			$\Upsilon_{t3} (2 ^3D_3) $     &  $\chi_{t2}(2 ^3P_2)\gamma$   & 148.968&0.92622&63.38   \\
			&  $\chi_{t2}(1 ^3P_2)\gamma$   &  482.662&0.52674 &36.04 \\
			&  $\chi_{t2} (1 ^3F_2)\gamma$   & 21.9993&0.000026 &0.0177\\
			&  $\chi_{t3} (1 ^3F_3)\gamma$   & 21.9993&0.000634 & 0.04338\\
			&  $\chi_{t4} (1 ^3F_4)\gamma$   &21.9993&0.007330 & 0.50164\\
			&Total&   &1.4611   &100\\
			\hline
		\end{tabular}
		\caption{Radiative widths and branching ratios for 2D toponium mesons.}\label{results-8}
	\end{table}
	\begin{table}[h!]
		\centering
		\begin{tabular}{c c c c c c}
			\hline
			\hline
			Initial State&Final State&$E_\gamma $ (MeV)&Predicted Width(KeV)&Predicted B.R($\%$)\\
			\hline\hline
			$\eta_{t3} (3 ^1D_2) $      &  $h_t(3 ^1P_1)\gamma$   &100.985&0.54425&12.58 \\
			&  $h_t(2 ^1P_1)\gamma$   &267.896&0.50478&11.67\\
			&  $h_t(1 ^1P_1)\gamma$   &601.474&3.25212 & 75.18\\
			&  $h_{t3} (2 ^1F_3)\gamma$   & 21.9993&0.019466&0.4499  \\
			&  $h_{t3} (1 ^1F_3)\gamma$   &  139.972&0.005364& 0.1239\\
			&  $\Upsilon_{t1}(1 ^3 D_1)\gamma$   &285.881&$6.16\times10^{-9}$ &$1.42\times 10^{-7}$   \\
			&  $\Upsilon_{t2}(1 ^3 D_2)\gamma$   & 284.882&$1.01\times10^{-8}$&$2.33\times 10^{-7}$   \\
			&  $\Upsilon_{t3}(1 ^3 D_3)\gamma$   &  284.882&$1.42\times10^{-8}$& $3.28\times 10^{-7}$ \\
			&Total&  &4.3259  &100\\
			\hline
			$\Upsilon_{t1} (3 ^3D_1) $     &  $\chi_{t0}(3 ^3P_0)\gamma$   & 99.9855&0.29413 &9.80  \\
			&  $\chi_{t1}(3 ^3P_1)\gamma$   & 99.9855&0.22060 & 7.35\\
			&  $\chi_{t2}(3 ^3P_2)\gamma$   & 98.9858&0.01427 & 0.475\\
			&  $\chi_{t0}(2 ^3P_0)\gamma$   & 267.896&0.28144 & 9.37\\
			&  $\chi_{t1}(2 ^3P_1)\gamma$   & 266.897&0.20873 &6.95\\
			&  $\chi_{t2}(2 ^3P_2)\gamma$   &  138.972&0.00363& 0.12\\
			&  $\chi_{t0}(1 ^3P_0)\gamma$   & 604.47&1.09507 &36.48\\
			&  $\chi_{t1}(1 ^3P_1)\gamma$   & 601.475&0.80916 & 26.96 \\
			&  $\chi_{t2}(1 ^3P_2)\gamma$   & 599.478&0.05340 & 1.77\\
			&  $\eta_{t2} (1 ^1D_2)\gamma$   &116.98&$5.90\times10^{-11}$ &$1.93\times 10^{-9}$  \\
			&  $\chi_{t2} (2 ^3F_2)\gamma$   &20.9994&0.01685 &  0.56\\
			&  $\chi_{t2} (1 ^3F_2)\gamma$   &138.972&0.00362& 0.12\\
			&Total&  &3.00   &100\\
			\hline
			$\Upsilon_{t2} (3 ^3D_2) $     &  $\chi_{t1}(3 ^3P_1)\gamma$   &99.9855&0.39709&12.49   \\
			&  $\chi_{t2}(3 ^3P_2)\gamma$   & 98.9858&0.12843& 4.04 \\
			&  $\chi_{t1}(2 ^3P_1)\gamma$   & 266.897&0.37571 & 11.82\\
			&  $\chi_{t2}(2 ^3P_2)\gamma$   &  266.897&0.12383& 3.895 \\
			&  $\chi_{t1}(1 ^3P_1)\gamma$   &  601.475&1.4564 & 45.81\\
			&  $\chi_{t2}(1 ^3P_2)\gamma$   &    599.478&0.48068&15.12\\
			&  $\eta_{t2}(2 ^1D_2)\gamma$   &  116.98&$8.25\times10^{-16}$ &$2.595\times 10^{-14}$ \\
			&  $\eta_{t2}(1 ^1D_2)\gamma$   &  116.98&$5.90\times10^{-11}$&$5.9 \times 10^{-11}$\\
			&  $\chi_{t2} (2 ^3F_2)\gamma$   & 20.9994&0.001872 & 0.0588\\
			&  $\chi_{t3} (2 ^3F_3)\gamma$   & 50.9962&0.21456&6.750\\
			&Total&  &3.1785 &100\\
			\hline
			$\Upsilon_{t3} (3 ^3D_3) $     &  $\chi_{t2}(3 ^3P_2)\gamma$   & 99.9855&0.52945 &17.65 \\
			&  $\chi_{t2}(2 ^3P_2)\gamma$   & 266.897&0.500952 & 16.70\\
			&  $\chi_{t2}(1 ^3P_2)\gamma$   &  600.477&1.93235 &64.44\\
			&  $\eta_{t2}(2 ^1D_2)\gamma$   &117.98&$8.46\times10^{-16}$&$2.82\times 10^{-14}$   \\
			&  $\eta_{t2}(1 ^1D_2)\gamma$   & 117.98&$6.05\times10^{-11}$&$2.01\times 10^{-19}$ \\
			&  $\chi_{t2} (2 ^3F_2)\gamma$   & 21.9993&0.0000439 & 0.00146\\
			&  $\chi_{t3} (2 ^3F_3)\gamma$   & 51.9961&0.020307 & 0.6772\\
			&  $\chi_{t4} (2 ^3F_4)\gamma$   &  20.9994&0.01548&0.5162\\
			&Total&   &2.998  &100\\
			\hline
		\end{tabular}
		\caption{Radiative widths and branching ratios for 3D toponium mesons.}\label{results-9}
	\end{table}
	\begin{table}[h!]
		\centering
		\begin{tabular}{c c c c c c}
			\hline
			\hline
			Initial State&Final State&$E_\gamma $ (MeV)&Predicted Width(KeV)&Predicted B.R($\%$)\\
			\hline\hline
			$\chi_{t2} (1 ^3F_2) $      &  $\Upsilon_{t1}(1 ^3D_1)\gamma$   &  145.969&1.18486& 84.27  \\
			&  $\Upsilon_{t2}(1 ^3D_2)\gamma$   & 144.97&0.21494  & 15.28\\
			&  $\Upsilon_{t3}(1 ^3D_3)\gamma$   & 144.97&0.006141 &0.4367\\
			&Total&   & 1.4059  &100\\
			\hline
			$\chi_{t3} (1 ^3F_3) $      &  $\Upsilon_{t2}(1 ^3D_2)\gamma$   & 144.97&1.22824 &88.89  \\
			&  $\Upsilon_{t3}(1 ^3D_3)\gamma$   &  144.97&0.15353 &11.11 \\
			&Total&   & 1.3817 & 100\\
			\hline
			$\chi_{t4} (1 ^3F_4) $      &  $\Upsilon_{t3}(1 ^3D_3)\gamma$   & 144.97&1.38177 &100  \\
			\hline
		\end{tabular}
		\caption{Radiative widths and branching ratios for 1F toponium mesons.}\label{results-11}
	\end{table}
	\section{Results and discussion}
	\label{results and discussion}
	In this work, numerical wave functions, mass spectrum, and root mean square radii of toponium mesons are computed with the non-relativistic quark potential model for 1S- 6S, 1P- 5P, 1D-4D, and 1F-3F states. Variation of numerical solutions (or wave functions) of $t \overline{t}$ with respect to $r$ is shown in Figs.(1-3) for different values of $L=0, 1, 2$. Each figure consist of four panels correspond to $n = 1, 2, 3, 4$, and each panel is associated with different values of $s, J$ for each value of $L$ and $n$. From these figures, it is observed that
	(i). number of nodes for each curve are equal to the quantum number $n-1$, and
	(ii). with the increase of $L$, peaks are moving away from the origin,
	(iii). no significant change in the wave functions for different values of $S$ and $J$ without changing $L$. This may be due to the heavy mass of toponium quark. As $\frac{1}{m^2_t}$ factor becomes very small ($\approx$ zero) that exists in the terms involving spin. Similar behaviour of wave functions for charmonium and bottomonium and $B_c$ are observed in refs. \cite{Nosheen14, Nosheen17, Nosheen19}.
	
	Mass spectrum and radii of toponium mesons are reported in Tables (1,2). Our mass prediction is in agreement with the recently observed mass of the $\eta_t$ meson at CERN which is reported as $m_t < 360$ GeV \cite{CMS-2025}. It is also observed that the calculated masses are very close to the mass spectrum reported in Ref.\cite{2025,2024a}. From Tables (1,2), it is observed that the masses and radii are increasing toward higher radial or orbital excitations. The increase in RMS radii with increase of orbital quantum number $L$ can be justified quantum mechanically. In Quantum Mechanics, centrifugal barrier increases with the increase of $L$. This fact weakens the binding strength between particles and results as an increase in the RMS radii of particles.
	
	Radiative transition widths are reported in Tables(\ref{results-1}-\ref{results-11}). Results show that M1 radiative widths are $\approx$ zero because of the factor $(\frac{1}{m^2_t})$ in the expression of M1 radiative widths. E1 radiative widths do not depend on factor $(\frac{1}{m^2_t})$, therefore this expression give higher values (of E1 radiative width) up to 25.32 KeV. As M1 widths are very small values, therefore branching ratios for E1 transitions are greater as compare to branching ratios for M1 transitions. In case of $3S \rightarrow 2P$, $2 D \rightarrow 1F$, E1 transitions are very small of order of $10^{-3}$ KeV, while E1 width is almost zero for $3S \rightarrow 1P$.
	
	\section{Concluding Remarks}
	\label{conclusions}
	In this paper, the spectrum, RMS radii, radiative transitions, and branching ratios of ground and higher states of toponium mesons are reported. Predictions presented in this work will help experimentalists to find the excited $t\overline{t}$ states and to measure their properties.
	
	\subsection*{Data Availability Statement}
	No Data associated in the manuscript.
	
	\subsection*{Funding Statement}
	No funding is available.

\end{document}